\documentclass{article}

\usepackage{arxiv}

\usepackage[utf8]{inputenc} % allow utf-8 input
\usepackage[T1]{fontenc}    % use 8-bit T1 fonts
\usepackage{hyperref}       % hyperlinks
\usepackage{url}            % simple URL typesetting
\usepackage{booktabs}       % professional-quality tables
\usepackage{amsfonts}       % blackboard math symbols
\usepackage{nicefrac}       % compact symbols for 1/2, etc.
\usepackage{microtype}      % microtypography
\usepackage{lipsum}		% Can be removed after putting your text content
\usepackage{graphicx}
\usepackage{doi}
\usepackage{stfloats}
\usepackage{diagbox}
\usepackage{xcolor}
\usepackage{amsmath,amsfonts, amssymb}
\usepackage[caption=false]{subfig} 
\usepackage{balance}%, subfigure
\usepackage[shortcuts,acronym]{glossaries}

\renewcommand{\arraystretch}{1.3}

\renewcommand*{\glossaryentrynumbers}[1]{}
\newacronym{ai}{AI}{artificial intelligence}
\newacronym{ml}{ML}{machine learning}
\newacronym{dl}{DL}{deep learning}
\newacronym{rl}{RL}{reinforcement learning}
\newacronym{dnn}{DNN}{deep neural network}
\newacronym{cnn}{CNN}{convolutional neural network}
\newacronym{rnn}{RNN}{recurrent neural network}
\newacronym{lstm}{LSTM}{long short term Memory}
\newacronym{gan}{GAN}{generative adversarial network}

\newacronym{vlc}{VLC}{visible light communication}
\newacronym{owc}{OWC}{optical wireless communication}
\newacronym{uwoc}{UWOC}{underwater wireless optical communication}
\newacronym{uwsn}{UWSN}{underwater wireless sensors network}
\newacronym{uwc}{UWC}{underwater wireless communication}
\newacronym{uac}{UAC}{underwater acoustic communication}
\newacronym{auv}{AUV}{autonomous underwater vehicles}
\newacronym{w2a}{W2A}{water-to-air}
\newacronym{a2w}{A2W}{air-to-water}
\newacronym{led}{LED}{light-emitting diode}
\newacronym{ld}{LD}{laser diode}
\newacronym{apd}{APD}{avalanche photodiode}
\newacronym{spad}{SPAD}{single-photon avalanche diode}
\newacronym{sipm}{SiPM}{silicon photo-multiplier}
\newacronym{imdd}{IM/DD}{intensity modulation direct-detection}
\newacronym{ook}{OOK}{on-off keying}
\newacronym{pam}{PAM}{pulse amplitude modulation}
\newacronym{qam}{QAM}{quadrature amplitude modulation}
\newacronym{qpsk}{QPSK}{quadrature phase shift keying}
\newacronym{ofdm}{OFDM}{orthogonal frequency division multiplexing}
\newacronym{oam}{OAM}{orbital angular momentum}
\newacronym{snr}{SNR}{signal-to-noise ratio}
\newacronym{ber}{BER}{bit error rate}
\newacronym{ser}{SER}{symbol error rate}
\newacronym{asv}{ASV}{autonomous surface vehicle}
\newacronym{mc}{MC}{Monte-Carlo}
\newacronym{pe}{PE}{pointing error}
\newacronym{nrzook}{NRZ-OOK}{non-return-to-zero on-off keying}
\newacronym{tia}{TIA}{trans-impedance amplifier}
\newacronym{lpf}{LPF}{low-pass filter}
\newacronym{pdf}{PDF}{probability density function}
\newacronym{pd}{PD}{photodetector}
\newacronym{hn}{HN}{Hall-Novarini}
\newacronym{aoa}{AoA}{angle-of-arrival}
\newacronym{fov}{FoV}{field-of-view}

\newacronym{iout}{IoUT}{internet of underwater things}
\newacronym{gbps}{Gbps}{gigabits per second}
\newacronym{kbps}{Kbps}{kilobits per second}
\newacronym{tx}{Tx}{transmitter}
\newacronym{rx}{Rx}{receiver}
\newacronym{leo}{LEO}{low earth orbit}
\newacronym{uav}{UAV}{unmanned aerial vehicle}
\newacronym{sk}{SK}{shift-keying}
\newacronym{fso}{FSO}{free space optics}
\newacronym{lg}{LG}{Laguerre-Gaussian}
\newacronym{em}{EM}{expectation maximization}
\newacronym{mse}{MSE}{mean squared error}
\newacronym{bmm}{BMM}{Beta mixture model}
\newacronym{tir}{TIR}{total internal reflection}
\newacronym{cdf}{CDF}{cumulative distribution function}
\newacronym{pat}{PAT}{pointing, acquisition, and tracking}

\newacronym{kl}{KL}{Kullback-Leibler}
\newacronym{ks}{KS}{Kolmogorov-Smirnov}
\newacronym{bic}{BIC}{bayesian information criterion}				%% fichiers sources glossaires
\setglossarystyle{super}

\title{Statistical Characterization of Wind-Induced Beam Refraction and UAV Instability in Water-to-Air Optical Channels}

\author{M. Nennouche\\
	Aix-Marseille University, CNRS\\
	Centrale Med, LIS\\
	Marseille, France \\
	\texttt{mohamed.nennouche@centrale-med.fr} \\
	\And
	I. C. Ijeh \\
	Department of Electrical/Electronic Engineering\\
	AE-FUNAI\\
	Ebonyi State, Nigeria \\
	\texttt{ikenna.ijeh@funai.edu.ng} \\
    \And
	M. A. Khalighi \\
	Aix-Marseille University, CNRS\\
	Centrale Med, Fresnel Institute\\
	Marseille, France \\
	\texttt{ali.khalighi@fresnel.fr} \\
}

\renewcommand{\shorttitle}{Nennouche \textit{et al.} (2026)}

\hypersetup{
pdftitle={A template for the arxiv style},
pdfsubject={q-bio.NC, q-bio.QM},
pdfauthor={David S.~Hippocampus, Elias D.~Striatum},
pdfkeywords={First keyword, Second keyword, More},
}

\begin{document}
\maketitle

\begin{abstract}
	Direct water-to-air (W2A) optical communications experience strong beam refraction at the dynamic sea surface and unmanned aerial vehicle (UAV) instability. This letter proposes a novel and tractable statistical channel model for a vertical W2A link between an underwater node and an UAV under varying wind speeds, modeling wind-induced pointing errors with a Beta mixture fitted via the Expectation–Maximization algorithm. By accounting for link interruptions due to total internal reflection (TIR) and effective receiver field-of-view limitations, we derive closed-form expressions for the channel distribution and link outage probability. Our analysis reveals a fundamental TIR-induced outage floor limiting link reliability and providing insight for robust W2A system design.
\end{abstract}

% keywords can be removed

\section{Introduction}\label{sec:introduction}

Underwater communications are attracting growing interest, driven by the rapid expansion of the \gls{iout} across diverse applications~\cite{Kaushal-Access-2016, Luo-Sensors-2022}.   %, ranging from industrial pipeline inspection to environmental coral reef monitoring
These use-cases require data rates and bandwidths beyond the capabilities of conventional acoustic systems. \glspl{uwoc} have thus emerged as a compelling alternative, enabling high-speed data transmission over distances of several tens of meters \cite{Kaushal-Access-2016}.
In \gls{uwoc} systems, retrieving data to the surface for remote processing remains a key challenge. The most common approach relies on surface platforms, such as buoys or autonomous surface vehicles \cite{Luo-Sensors-2022}, introducing additional cost and environmental impact. As a result, direct \gls{w2a} communication has recently gained significant traction \cite{Luo-Sensors-2022}. This approach, however, introduces increased channel complexity due to the dynamic nature of the sea surface. Wave-induced random refraction and reflection \cite{Luo-Sensors-2022, Nennouche-Photonics-2026} make accurate modeling of the direct \gls{w2a} channel highly challenging, yet such modeling is essential for ensuring reliable communication.

Several recent studies have focused on modeling the direct \gls{w2a} channel, primarily through \gls{mc} simulations, developing realistic simulators that incorporate various sea surface models and underwater impairments, including absorption and scattering by suspended particles, as well as the effects of air bubbles generated by wave breaking \cite{Nennouche-Photonics-2026, Lin-JLT-2022, Angara-OLT-2024}. \textcolor{black}{Recent works have also proposed statistical channel models for \gls{w2a} \cite{Ata-TCOM-2024} and \gls{a2w} \cite{Rahman-TVT-2022} links. However,
the ray-tracing simulator in \cite{Nennouche-Photonics-2026} is computationally prohibitive for system-level design, while the statistical models in \cite{Ata-TCOM-2024, Rahman-TVT-2022} remain decoupled from wind-driven surface dynamics and overlook link interruptions caused by \gls{tir} and \gls{uav} instability.
%accurate, the ray-tracing simulator of \cite{Nennouche-Photonics-2026} entails a prohibitive computational cost for system-level design, whereas the statistical models of \cite{Ata-TCOM-2024, Rahman-TVT-2022} remain decoupled from the wind dynamics driving the surface roughness and overlook the two mechanisms that break the link altogether, namely \gls{tir} and \gls{uav} aerodynamic instability. 
This letter bridges this gap by proposing a comprehensive wind-dependent statistical framework that preserves the physical fidelity of \cite{Nennouche-Photonics-2026} within a closed-form formulation, enabling computationally efficient 
training of advanced pointing and tracking schemes \cite{Wang-TVT-2024}.}
%environment for training advanced pointing and tracking schemes \cite{Wang-TVT-2024}.}
\textcolor{black}{Specifically, we model sea surface beam refraction and \gls{uav} tilt as functions of wind speed and derive a tractable closed-form expression for the \gls{pe} loss based on a Beta mixture distribution, together with the link outage probability accounting for \gls{tir} and \gls{fov} limitations.} Our analytical results are validated through extensive \gls{mc} simulations, enabling a detailed analysis of the outage performance.

%%%%%%%%%%%%%%%%%%%%%%%%%%%%%%%%%%%%%%%%%%%%%%%%%%%%%%%%%%%%%%%%%%%%%%%%%%%%%
\vspace{-0.3cm}
\section{Link Geometry and System Model}
\label{sec:system_model}

\begin{figure}[!t]
\centering
\includegraphics[width=0.475\textwidth]{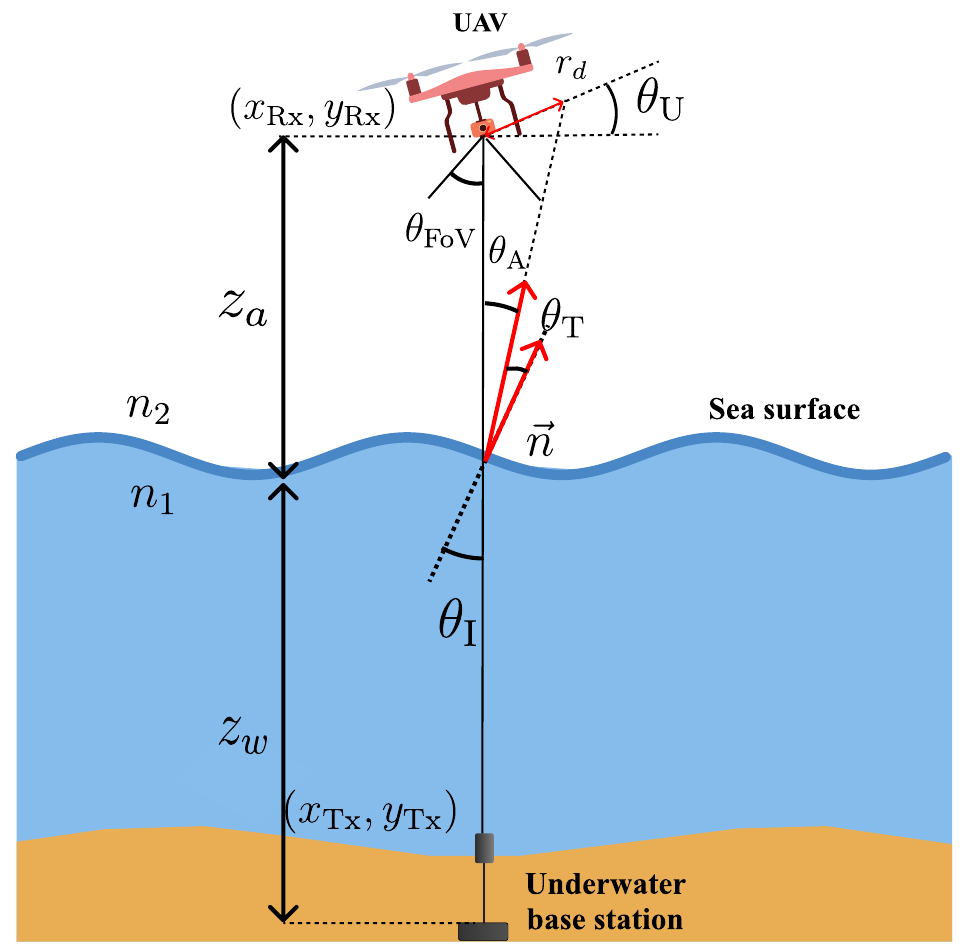}
\caption{\textcolor{black}{Illustration of the vertical water-to-air (W2A) optical wireless communication system model.}}
\label{fig:systemmodel}
\end{figure}

We consider a vertical \gls{w2a} link between a base station equipped with a \gls{ld} at depth $z_w$, and an \gls{uav} hovering at altitude $z_a$ above the sea surface, equipped with a \gls{pd} with \gls{fov} $\theta_\mathrm{FoV}$. Perfect alignment between the \gls{tx} and \gls{rx} is assumed, as illustrated in Fig.~\ref{fig:systemmodel}. We consider a dynamic sea surface, with an incident angle $\theta_\mathrm{I}$ defined between the surface normal $\vec{n}$ and the \gls{ld} beam, which is refracted to an angle $\theta_\mathrm{A}$ with respect to the vertical axis, \textcolor{black}{while the \gls{uav} experiences a wind-induced tilt angle, denoted by $\theta_\mathrm{U}$ \cite{Nennouche-Photonics-2026}.}%aerodynamic
%........................................
%........................................
We employ \gls{nrzook} modulation, where transmit powers $P_{\mathrm{Tx}, 1}$ and $P_{\mathrm{Tx}, 0} = \xi P_{\mathrm{Tx},1}$ are assigned to bits `1' and `0', respectively,  with $\xi$ denoting the extinction ratio. At the \gls{rx}, a high-sensitivity \gls{sipm} is employed to extend the transmission range. %, \textcolor{red}{followed by a \gls{tia}, and a \gls{lpf} to limit the Rx noise level.}

%%%%%%%%%%%%%%%%%%%%%%%%%%%%%%%%%%%%%%%%%%%%%%%%%%%%%%%%%%%%%%%%%%%%%%%%%%%%%%%%%%%%%%%%%%%%%%%%%%%%%%%%%
\vspace{-0.3cm}
\section{\textcolor{black}{Channel Modeling}}

\textcolor{black}{The overall channel loss is expressed by $h=h_\mathrm{L}\, h_\mathrm{P}$, where $h_\mathrm{L}$ denotes the deterministic path loss and $h_\mathrm{P}$ the random \gls{pe}-induced loss. Oceanic turbulence-induced scintillation is neglected, as the large \gls{rx} aperture (i.e., the \gls{pd} lens) mitigates irradiance fluctuations through aperture averaging \cite{Ijeh-JOCN-2022}. Furthermore, given the short in-air propagation distance and broadened beam footprint over the vertical underwater segment, the effects of atmospheric turbulence and underwater beam wandering are considered negligible compared to with surface-induced refraction.}

\vspace{-0.3cm}
\subsection{\textcolor{black}{Path Loss}}

To formulate $h_\mathrm{L}$, we assume that atmospheric attenuation over the relatively short in-air segment is negligible, such that the overall path loss is dominated by the underwater segment. This attenuation is due to absorption, scattering, and wave-generated air bubbles \cite{Chen-JOSAA-2024}. Accordingly: 
\begin{equation}
h_\mathrm{L} = \exp\left( -\int_0^{z_w} c_w(\lambda, z)dz \right),
\end{equation}
where $\lambda$ and $z$ denote wavelength and depth, respectively, and $c_w(\lambda, z) = a(\lambda) + b(\lambda) + b_\mathrm{bub}(z, U)$ denotes the attenuation coefficient. Here, $a(\lambda)$, $b(\lambda)$, and $b_\mathrm{bub}(z, U)$ represent the absorption coefficient, the scattering coefficient, and the bubble-induced scattering coefficient, respectively \cite{Nennouche-Photonics-2026, Chen-JOSAA-2024}. To calculate $b_\mathrm{bub}$, we use the \gls{hn} model \cite{Nennouche-Photonics-2026, Chen-JOSAA-2024}:
\begin{equation}
b_\mathrm{bub} = Q_\mathrm{sca}\Psi~N_b(z),
\end{equation}
where $Q_\mathrm{sca} = 2.0$ and $\Psi \approx 3\pi r_\mathrm{min}^2$ represent the mean scattering efficiency and the mean geometric cross-sectional area of the bubble population, respectively. Also, $N_b(z)$ represents the bubble density at depth $z$, approximated as \cite{Nennouche-Photonics-2026}:
\begin{equation}
N_b(z) \approx (1.6\times10^{10})\frac{r_{\mathrm{ref}}^4}{3\,r_{\mathrm{min}}^3}\bigg(\frac{U}{13}\bigg)^3 \exp\left[-\frac{z}{L(U)}\right].
\end{equation}
Here, $r_\mathrm{min} = 1\,\mu\mathrm{m}$ and $r_\mathrm{ref} = 54.4\,\mu\mathrm{m} + 1.984\times 10^{-6}z$ denote the minimum and reference air bubble radii, respectively. Lastly, $L(U)$ represents the e-folding distance, defined as:
\begin{equation}
L(U)= \begin{cases}
0.4, & U \leq 7.5\ \mathrm{m}/\mathrm{s}, \\
0.4+0.115\left(U-7.5\right), & U>7.5\ \mathrm{m}/\mathrm{s}.
\end{cases}
\end{equation}

\subsection{\textcolor{black}{Pointing Error Loss}}

Assuming a static \gls{tx} and \textcolor{black}{a hovering \gls{uav} experiencing a wind-induced tilt angle $\theta_\mathrm{U}$, the \glspl{pe} stem from both the random motion of the sea surface and the aerodynamic deviation of the \gls{uav} \cite{Nennouche-Photonics-2026}.} This results in fluctuations in the radial displacement $r_d$ at the \gls{rx} plane. \textcolor{black}{The resulting \gls{pe} loss $h_\mathrm{P}$ is characterized by two distinct regimes: if the optical beam is successfully captured by the \gls{rx}, the loss follows the geometric model of \cite{Ijeh-JOCN-2022}; otherwise, a complete link interruption occurs, yielding $h_\mathrm{P} = 0$. This conditional behavior is formulated as follows:
\begin{equation}
h_\mathrm{P} = \begin{cases}
A_0 \exp\left(-\dfrac{2r_d^2}{\omega_\mathrm{Leq}^2}\right) &
\text{if } (\theta_\mathrm{I} \le \theta_c) \text{ and } (\psi_\mathrm{A} \le \theta_\mathrm{FoV}), \\[2mm]
0 & \text{otherwise},
\end{cases}
\label{eq:hpphysical}
\end{equation}
where $A_0 = \left( \mathrm{erf}(v) \right)^2$ represents the maximum fraction of received power, with $\mathrm{erf}(\cdot)$ denoting the error function and $v = \left( D_r\sqrt{\pi} \right)/ \left( 2\sqrt{2}~\omega_L \right)$. Here, $D_r$ is the \gls{pd} lens diameter, $\omega_L$ is the beam waist at the total propagation distance \mbox{$L = z_w + z_a$}, and $\omega_\mathrm{Leq}= \omega_L \sqrt{\sqrt{\pi}~\mathrm{erf}(v) /\left[ 2v\exp\left(-v^2\right) \right]}$ denotes the equivalent beam width at the \gls{rx}.}
\textcolor{black}{The condition $\theta_\mathrm{I} \leq \theta_c$ ensures that the incidence angle does not exceed the critical angle $\theta_c$ defined by Snell's law, thereby avoiding \gls{tir}. Also, the condition $\psi_\mathrm{A} \leq \theta_\mathrm{FoV}$ dictates that the effective \gls{aoa} $\psi_\mathrm{A} = \theta_\mathrm{A} + \theta_\mathrm{U}$ falls within the \gls{rx} \gls{fov}.}

Given that the beam deviation occurs at the surface, assuming $x_\mathrm{Tx} = y_\mathrm{Tx} = x_\mathrm{Rx} = y_\mathrm{Rx} = 0$, \textcolor{black}{and accounting for the \gls{uav} tilt angle $\theta_\mathrm{U}$, the radial displacement is geometrically derived as
\begin{equation}
  r_d = z_a \frac{\sin(\theta_\mathrm{A})}{\cos(\psi_\mathrm{A})}.
  \label{eq:expr_rd}
\end{equation}}

\textcolor{black}{Note that the wave-induced refraction and the \gls{uav} deviation are assumed coplanar and oppositely oriented, with an isotropic sea-surface slope. Both assumptions overestimate $r_d$, yielding a worst-case model that lower-bounds the achievable link performance. Characterizing the \gls{pe} loss thus requires the statistics of $\psi_\mathrm{A}$ and the probability of violating the capture condition in \eqref{eq:hpphysical}, addressed in the following two subsections.}

\vspace{-0.2cm}
\subsection{\textcolor{black}{AoA Statistical Modeling}}\label{sec:aoa}

\textcolor{black}{Let $\theta_\mathrm{T}$ denote the transmission angle relative to the surface normal $\vec{n}$ (see Fig.~\ref{fig:systemmodel}). As introduced above, the effective \gls{aoa} is defined as $\psi_\mathrm{A} = \theta_\mathrm{A} + \theta_\mathrm{U}$, where $\theta_\mathrm{A} = \theta_\mathrm{T} - \theta_\mathrm{I}$ is the absolute deviation angle of the refracted beam and $\theta_\mathrm{U}$ is the wind-induced \gls{uav} tilt angle, modeled by a Gamma distribution \cite{Nennouche-Photonics-2026}. The angle $\theta_\mathrm{T}$ is obtained via Snell's law as $\theta_\mathrm{T} = \arcsin\left(\frac{n_1}{n_2}\sin(\theta_\mathrm{I})\right)$, where $n_1 \approx 1.33$ and $n_2 \approx 1$ denote the refractive indices of water and air, respectively \cite{OceanWb-Mobley}.}

\textcolor{black}{Modeling $\theta_\mathrm{I}$ requires characterizing the spatio-temporal dynamics of the sea surface, which are governed by environmental factors such as the wind speed $U$ and remain an active research topic \cite{Ata-TCOM-2024, Ijeh-CSNDSP-2026}. We adopt here the model of \cite{Ijeh-CSNDSP-2026}, where $\theta_\mathrm{I}$ follows a Weibull distribution whose \gls{pdf} is given by:}
\begin{equation}\label{eq:theta-I}
f_{\theta_\mathrm{I}}(\theta_\mathrm{I}) = \frac{k_U}{\lambda_U}\left( \frac{\theta_\mathrm{I}}{\lambda_U} \right)^{k_U-1}\exp\left[ -\left( \frac{\theta_\mathrm{I}}{\lambda_U} \right)^{k_U} \right],
\end{equation}
where $k_U = 1.7454 + 0.0071\,U$ and $\lambda_U = 13.6485 + 0.2406\,U$ are the \gls{pdf} shape and scale parameters, respectively, valid for wind speeds $U \in [6, 15]$ m/s.

Due to the highly non-linear nature of this geometric transformation, deriving tractable, closed-form \glspl{pdf} for \textcolor{black}{$\theta_\mathrm{A}$ and $\psi_\mathrm{A}$} is mathematically prohibitive. Therefore, we adopt a robust statistical approximation based on random numerical sampling. \textcolor{black}{For each wind speed, both angles were fitted to the same set of candidate distributions, namely Rayleigh,
Nakagami, Weibull, Rice, and Gamma, and evaluated using the goodness-of-fit metrics of \gls{mse}, $R^2$, \gls{kl} divergence, \gls{ks} statistic, and \gls{bic}. The results indicate that the effective \gls{aoa} $\psi_\mathrm{A}$ is best approximated by a Gamma distribution with shape and scale parameters $\alpha_\mathrm{A}$ and $\beta_\mathrm{A}$, and the deviation angle $\theta_\mathrm{A}$ likewise follows a Gamma distribution, with parameters denoted by $\alpha_\theta$ and $\beta_\theta$. This is consistent with the additive structure $\psi_\mathrm{A} = \theta_\mathrm{A} + \theta_\mathrm{U}$, where the Gamma-distributed \gls{uav} tilt shifts and broadens the refracted-beam statistics while preserving their distributional family.}

\begin{figure*}
\centering
\subfloat[$U = 6$ m/s]{\includegraphics[width=0.475\textwidth]{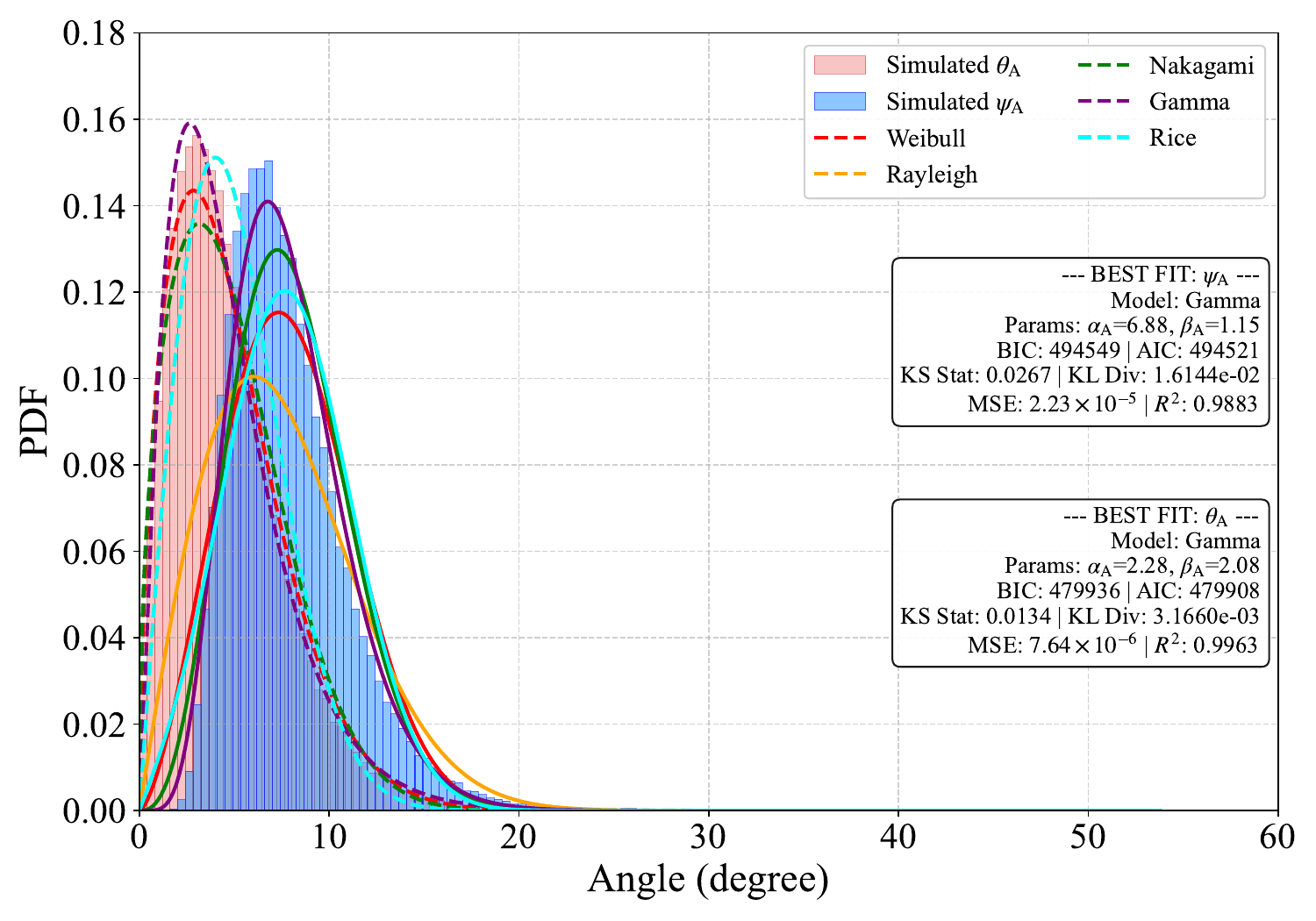}\label{fig:AoA_U_6}}
\hfill
\subfloat[$U = 14$ m/s]{\includegraphics[width=0.475\textwidth]{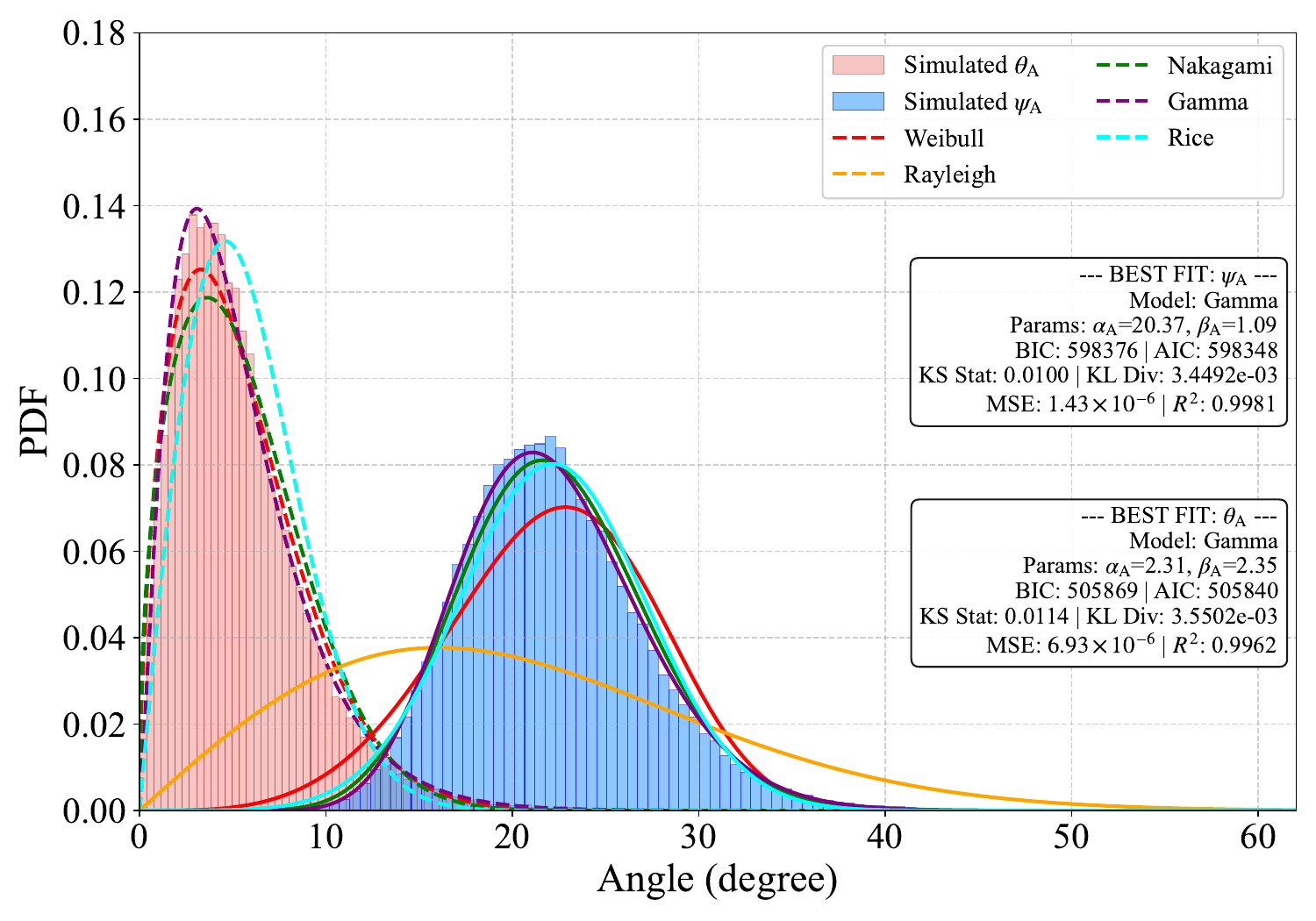}\label{fig:AoA_U_14}}\\
\subfloat[]{\includegraphics[width=0.55\textwidth]{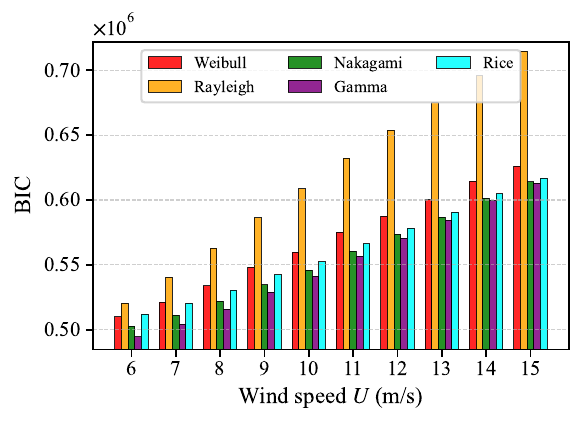}\label{fig:bic_barchart}}
\caption{\textcolor{black}{Statistical characterization of the deviation angle $\theta_\mathrm{A}$ and the effective \gls{aoa} $\psi_\mathrm{A}$ for (a) $U = 6$~m/s and (b) $U = 14$~m/s, and (c) \gls{bic} of the candidate models for $\psi_\mathrm{A}$ versus wind speed. Sampling is based on $2\times10^5$ independent trials per link configuration. Note that $\theta_\mathrm{A}$ is bounded by the critical angle $\theta_c = \arcsin\left(n_2/n_1\right) = 48.75^\circ$, beyond which \gls{tir} occurs.}}
\label{fig:AoA_fits}
\end{figure*}

\textcolor{black}{Figures~\ref{fig:AoA_U_6} and \ref{fig:AoA_U_14} reveal the cumulative impact of surface refraction and \gls{uav} instability: as $U$ increases from $6$ to $14$~m/s, the effective AoA broadens from approximately $0$--$20^\circ$ to $10$--$40^\circ$, resulting in substantially higher channel loss. Moreover, $\psi_\mathrm{A}$ is consistently shifted relative to $\theta_\mathrm{A}$ due to the \gls{uav} tilt. The suitability of the Gamma model for $\psi_\mathrm{A}$ is further supported by Fig.~\ref{fig:bic_barchart}, where the Gamma distribution consistently yields the lowest \gls{bic} over the entire wind-speed range, indicating the best fit among the considered models.}

\textcolor{black}{Finally, regression analysis over the considered wind-speed range yields $\alpha_\theta \approx 2.244$ and $\beta_\theta \approx 0.0365\,U + 1.878$, and $\alpha_\mathrm{A} \approx 1.68\,U - 3.11$ and $\beta_\mathrm{A} \approx 0.007\,U^2 - 0.144\,U + 1.75$ for $\psi_\mathrm{A}$. Thus, the proposed angular statistical model is fully parameterized by $U$ within the considered operating range.}

\vspace{-0.3cm}
\subsection{\textcolor{black}{Link Interruption Modeling}}

\textcolor{black}{It follows from \eqref{eq:hpphysical} that a complete link interruption occurs whenever the incidence angle exceeds the critical angle \mbox{($\theta_\mathrm{I} > \theta_c$)} or the effective \gls{aoa} falls outside the \gls{rx} \gls{fov} ($\psi_\mathrm{A} > \theta_\mathrm{FoV}$). The interruption probability $P_\mathrm{int}$ is then:
\begin{equation}
    P_\mathrm{int} = 1 - P(\theta_\mathrm{I} \le \theta_c \cap \psi_\mathrm{A} \le \theta_\mathrm{FoV}).
\end{equation}
These two events are not statistically independent, since the \gls{rx} can only perceive an angle $\psi_\mathrm{A}$ if the ray first survives \gls{tir} at the water-air interface. Using Bayes' theorem, the joint probability can be factorized as:
\begin{multline}
    P(\theta_\mathrm{I} \le \theta_c \cap \psi_\mathrm{A} \le \theta_\mathrm{FoV}) = P(\theta_\mathrm{I} \le \theta_c) \\ \times P(\psi_\mathrm{A} \le \theta_\mathrm{FoV} \mid \theta_\mathrm{I} \le \theta_c).
\end{multline}
Since the Gamma fit of $\psi_\mathrm{A}$ was performed exclusively on successfully refracted rays, it inherently represents this conditional probability. Hence, the above factorization is exact and does not rely on an independence assumption. Substituting the \glspl{cdf} of the Weibull and Gamma distributions then yields the following exact closed-form expression for the channel interruption probability:
\begin{equation}
    P_\mathrm{int} = 1 - \left[ \left( 1 - \exp\left(-\Big(\frac{\theta_c}{\lambda_U}\Big)^{k_U}\right) \right) \frac{\gamma\left(\alpha_\mathrm{A}, \frac{\theta_\mathrm{FoV}}{\beta_\mathrm{A}}\right)}{\Gamma(\alpha_\mathrm{A})} \right],
    \label{eq:pint}
\end{equation}
where $\gamma(\cdot, \cdot)$ and $\Gamma(\cdot)$ denote the lower incomplete Gamma function and the standard Gamma function, respectively. Note that the first factor depends solely on the sea-surface statistics, thereby imposing an irreducible \gls{tir}-induced floor on the link outage probability.}
\vspace{-0.2cm}
\subsection{\textcolor{black}{PE Distribution Modeling}}

\textcolor{black}{From \eqref{eq:hpphysical}, $h_\mathrm{P}$ is a mixed random variable whose \gls{pdf} is:
\begin{equation}
f_{h_\mathrm{P}}(h) = P_\mathrm{int}\, \delta(h) + (1 - P_\mathrm{int})\, f_{h_\mathrm{P},c}(h),
\label{eq:fhp_mixed}
\end{equation}
where $\delta(\cdot)$ is the Dirac delta function, accounting for link interruption events, and $f_{h_\mathrm{P},c}(\cdot)$ is the \gls{pdf} of $h_\mathrm{P}$ conditioned on successful beam capture by the \gls{rx}, given by \cite{Ijeh-JOCN-2022}:
\begin{equation}
f_{h_\mathrm{P},c}(h_\mathrm{P,c}) = \frac{\omega_\mathrm{Leq}^2}{4h_\mathrm{P,c}\sqrt{-\frac{\omega_\mathrm{Leq}^2}{2}\ln\left( \frac{h_\mathrm{P,c}}{A_0} \right)}} f_{r_d}\left( \sqrt{-\frac{\omega_\mathrm{Leq}^2}{2}\ln\left(\frac{h_\mathrm{P,c}}{A_0}\right)} \right),
\label{eq:f_h_p}
\end{equation}
where $f_{r_d}(\cdot)$ is the \gls{pdf} of $r_d$ in \eqref{eq:expr_rd}. Since \mbox{$\psi_\mathrm{A} = \theta_\mathrm{A} + \theta_\mathrm{U}$}, deriving $f_{r_d}(\cdot)$ requires the joint \gls{pdf} of $\theta_\mathrm{A}$ and $\psi_\mathrm{A}$, whose nonlinear coupling through trigonometric functions precludes a tractable closed-form expression for $f_{h_\mathrm{P},c}(\cdot)$ and, consequently, for key performance metrics such as the outage probability, which is critical for robust \gls{w2a} link design.} Here, we propose a novel statistical modeling approach based on a finite mixture model, optimized via the \gls{em} algorithm \cite{Moon-SPM-1996}. The \gls{em} algorithm iteratively estimates the latent component responsibilities (E-step) and updates the distribution parameters to maximize the log-likelihood (M-step).
To improve EM convergence and obtain a generalized model independent of the \gls{rx} aperture size, we apply EM to the normalized \gls{pe} loss \textcolor{black}{$h_{\mathrm{P, N}} = h_\mathrm{P, c}/A_0$.}
Based on extensive evaluations, we propose a \gls{bmm} to accurately characterize the \gls{pdf} of $h_{\mathrm{P, N}}$, expressed as:
\begin{equation}
f_{h_{\mathrm{P, N}}}(h_{\mathrm{P, N}}) \approx \sum_{i=1}^{2} w_i \frac{h_{\mathrm{P, N}}^{\alpha_i - 1} (1 - h_{\mathrm{P, N}})^{\beta_i - 1}}{\mathrm{B}(\alpha_i, \beta_i)}, \quad 0 \leq h_{\mathrm{P, N}} \leq 1,
\label{eq:f_p_n}
\end{equation}
where $w_i$ denotes the mixing weight of the $i$-th component (with $w_1+w_2=1$), and $\mathrm{B}(\cdot, \cdot)$ is the standard Beta function with $\alpha_i > 0$ and $\beta_i > 0$ as shape parameters.
\begin{figure}
\centering
\subfloat[$U = 6$ m/s]{\includegraphics[width=0.475\textwidth]{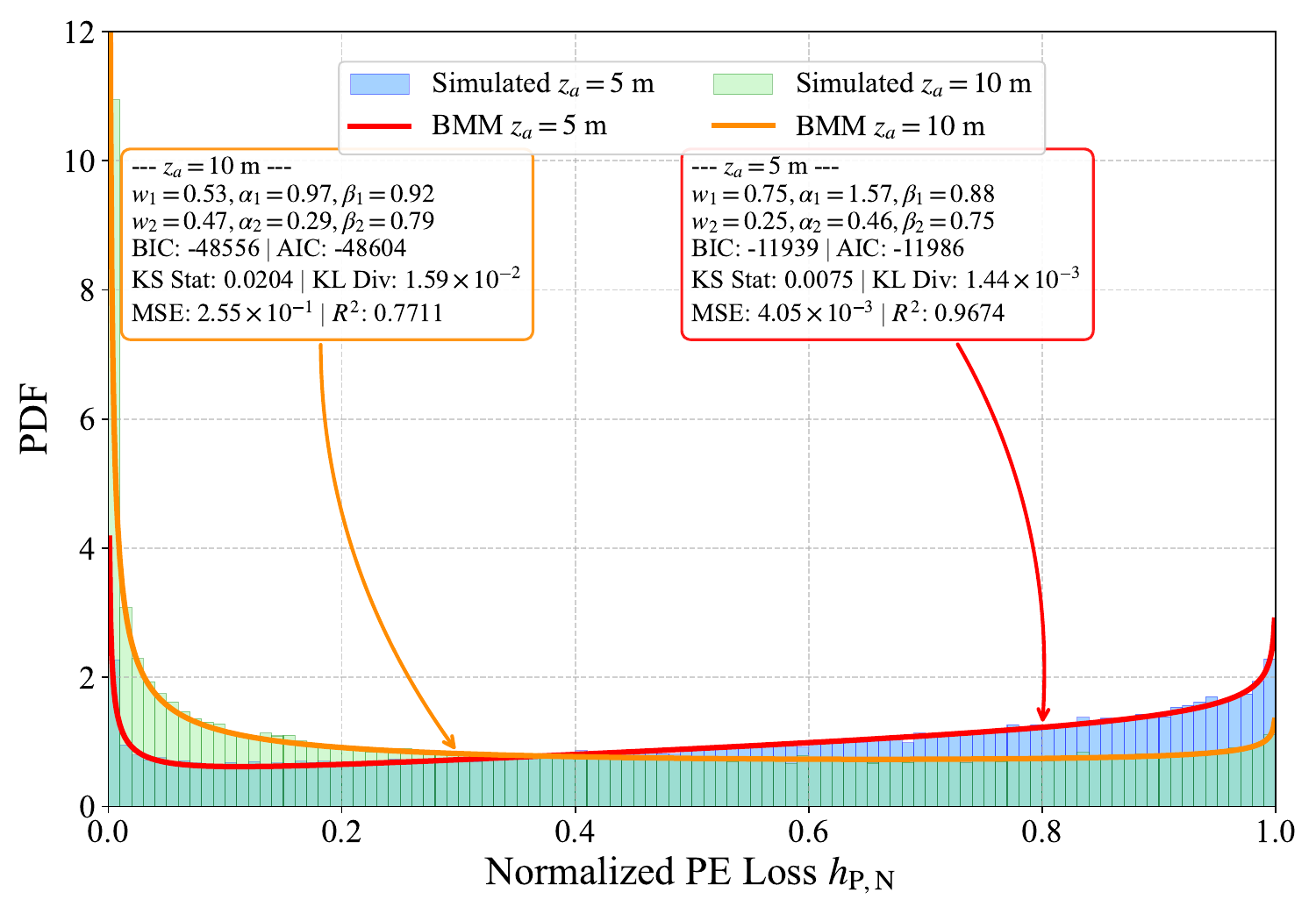}\label{fig:hp_EM_U_6}}
\hfill
\subfloat[$U = 14$ m/s]{\includegraphics[width=0.475\textwidth]{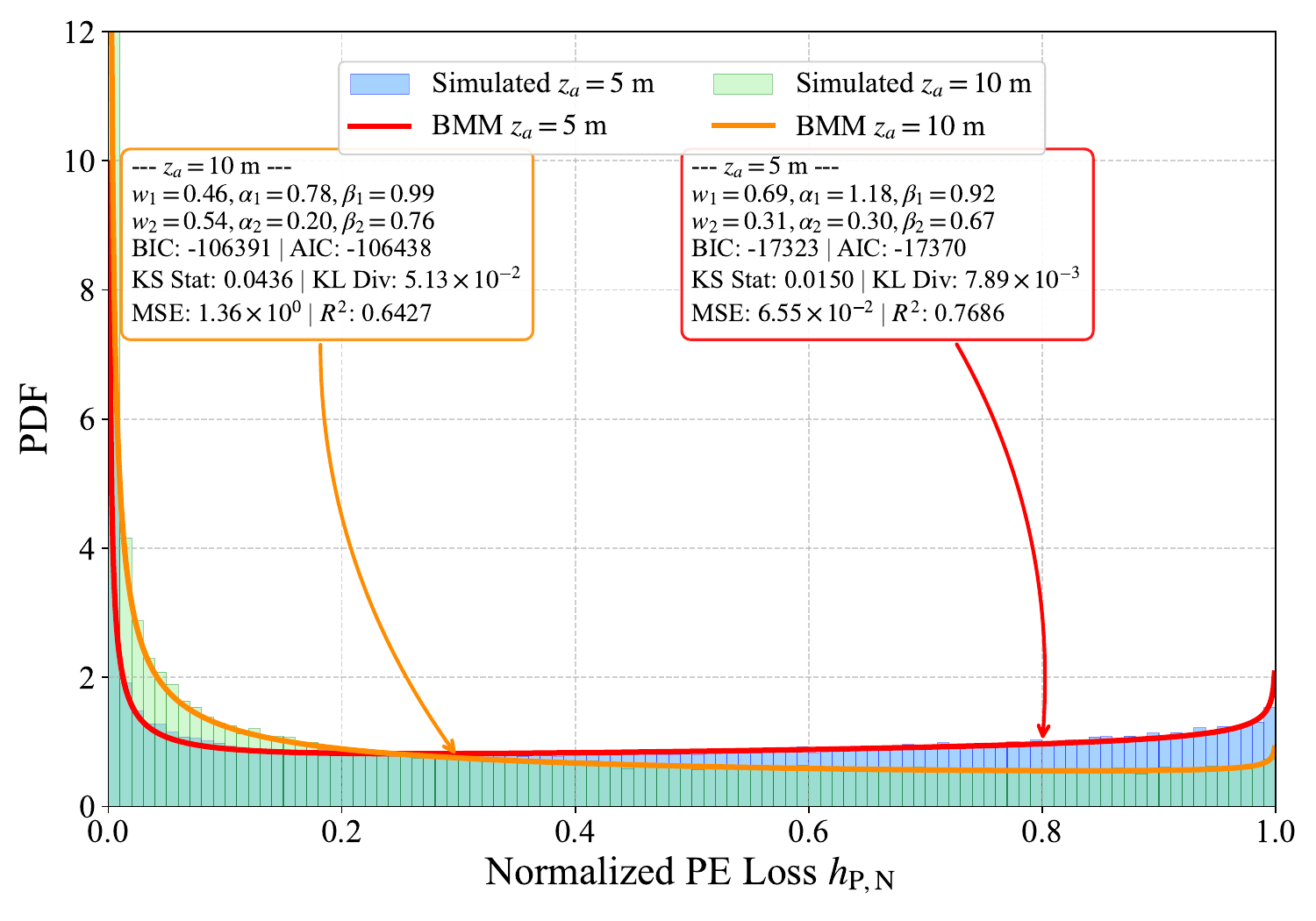}\label{fig:hp_EM_U_14}}
\caption{Comparison of the simulated exact physical pointing error distribution and the proposed EM-fitted Beta Mixture Model (BMM) for different wind speeds $U$ and air distance $z_a$.}
\label{fig:hp_EM_mixture}
\end{figure}

Figures~\ref{fig:hp_EM_U_6} \textcolor{black}{and \ref{fig:hp_EM_U_14}} show the simulated histograms of the normalized \gls{pe} loss alongside the proposed \gls{bmm} \gls{pdf} for wind speeds \textcolor{black}{$U = 6$ and $14$ m/s}. The system parameters are set to $z_w = 10$~m, $D_r = 75$~mm, and $\omega_L =L\tan(\theta_0)$, where $\theta_0=0.05$\,rad ($2.85^\circ$) is the \gls{ld} half divergence angle. The robustness of the \gls{bmm} is assessed for two \gls{uav} altitudes: $z_a = 5$~m and $10$~m. 
Although the \gls{pdf} shapes vary significantly with altitude, our proposed \gls{bmm} consistently adapts, providing a highly accurate fit in both scenarios. Furthermore, it is evident that at higher altitudes, the probability of severe fading ($h_\mathrm{P} \to 0$) increases substantially with wind speed.

Lastly, it is worth mentioning that, in \eqref{eq:f_p_n}, the summation is restricted to two components ($i=1,2$). \textcolor{black}{This choice reflects the bimodal nature of the distribution, with one component capturing the \gls{pdf} peak near $0$ and the other the peak near $1$ (see Fig.~\ref{fig:hp_EM_mixture}). It is further supported by evaluating mixtures of two to four components fitted over the entire wind-speed range and both \gls{uav} altitudes, using the same metrics in Section~\ref{sec:aoa}. The two-component model consistently yields the lowest \gls{bic}, (see Table~\ref{tab:bic_components}), thus offering the best accuracy--parsimony trade-off while preserving analytical tractability. Moreover, the fitted \gls{bmm} parameters evolve monotonically with the wind speed, $w_1$ and $\alpha_1$ decreasing while $w_2$ increasing, consistent with the progressive shift from stable hovering regime to deep-fade events. The \gls{bmm} thus reflects the underlying physical dynamics rather than merely fitting the empirical data. This motivates a full analytical mapping between the \gls{bmm} and physical link parameters, left for future work.}

\begin{table*}[t]

\centering
\caption{\textcolor{black}{\scriptsize BIC of the fitted BMM versus the number of components $M$ and the wind speed $U$, for $z_w = 10$~m and $z_a = 5$~m.}}
\label{tab:bic_components}
\setlength{\tabcolsep}{6pt}
\renewcommand{\arraystretch}{1}
\small
\begin{tabular}{c|cccccccccc}
\hline
\diagbox[width=7.2em,height=1.8\line]{$M$}{$U$ (m/s)} & 6 & 7 & 8 & 9 & 10 & 11 & 12 & 13 & 14 & 15 \\
\hline
2 & $\mathbf{-11939}$ & $\mathbf{-11057}$ & $\mathbf{-10670}$ & $\mathbf{-11066}$ & $\mathbf{-10367}$ & $\mathbf{-11419}$ & $\mathbf{-12749}$ & $\mathbf{-13570}$ & $\mathbf{-17323}$ & $\mathbf{-21401}$ \\
3 & $-11892$ & $-11010$ & $-10629$ & $-10994$ & $-10301$ & $-11275$ & $-12604$ & $-13396$ & $-17163$ & $-21183$ \\
4 & $-11830$ & $-10962$ & $-10550$ & $-10900$ & $-10226$ & $-11201$ & $-12527$ & $-13298$ & $-17071$ & $-21088$ \\
\hline
\end{tabular}

\end{table*}

%\vspace{-0.2cm}
\subsection{\textcolor{black}{Overall Channel Loss}}

As explained in Section~\ref{sec:system_model}, the overall channel coefficient is expressed as $h = h_\mathrm{L} h_\mathrm{P}$. \textcolor{black}{Since $h_\mathrm{P}$ is a mixed random variable, the \gls{pdf} of $h$ is also of mixed type:}
\begin{equation}
\label{eq:pdf_h_algebraic}
f_h(h) = P_\mathrm{int}\,\delta(h) + \left(1 - P_\mathrm{int}\right)f_c(h),
\end{equation}
where $f_c(h)$ is the \gls{pdf} of $h_c \triangleq h_\mathrm{L}h_\mathrm{P, c}$. \textcolor{black}{Using the normalized \gls{pe} model in \eqref{eq:f_p_n} and incorporating $h_\mathrm{L}$ and $A_0$ yields:}
\begin{equation}
f_{h_c}(x) = \frac{1}{h_\mathrm{L} A_0} \sum_{i=1}^{2} w_i \frac{\left(\frac{x}{h_\mathrm{L} A_0}\right)^{\alpha_i - 1} \left(1 - \frac{x}{h_\mathrm{L} A_0}\right)^{\beta_i - 1}}{\mathrm{B}(\alpha_i, \beta_i)},
\label{eq:f_c}
\end{equation}
which is valid for $0 \leq x \leq h_\mathrm{L} A_0$.
Substituting \eqref{eq:f_c} into \eqref{eq:pdf_h_algebraic}, and applying the identities in \cite[Eq.~(8.4.2.3)]{Prudnikov-book-1990} and \cite[Eq.~(8.2.2.15)]{Prudnikov-book-1990}, allows simplification of the fractional binomial terms. The exact closed-form \gls{pdf} of the overall channel $h$ for $0 \leq h \leq h_\mathrm{L} A_0$ is thus obtained as:
\begin{equation}
f_h(h) = P_\mathrm{int} \delta(h) + \frac{1 - P_\mathrm{int}}{h_\mathrm{L} A_0} \sum_{i=1}^{2} w_i \frac{\Gamma(\alpha_i + \beta_i)}{\Gamma(\alpha_i)} G_{1,1}^{1,0}\left( \frac{h}{h_\mathrm{L} A_0} \left| \begin{array}{c} \alpha_i + \beta_i - 1 \\ \alpha_i - 1 \end{array} \right.\right),
\label{eq:pdf_h_meijer}
\end{equation}
\textcolor{black}{where $\Gamma(\cdot)$ is the Gamma function and $G_{p,q}^{m,n}(\cdot|\cdot)$ the Meijer G-function \cite[Eq.~(8.2.1.1)]{Prudnikov-book-1990}.}

\section{Link Outage Performance}

A link outage occurs when the instantaneous \gls{ber} exceeds a target \gls{ber}.
%From \eqref{eq:snr}, 
This is equivalent to considering a threshold on the channel gain $h_\mathrm{th}$ \cite{Ijeh-JOCN-2022}, allowing the outage probability $P_\mathrm{out}$ to be obtained via the \gls{cdf} of $h$ as:
\begin{equation}
P_\mathrm{out}(h_\mathrm{th}) = \Pr(h \leq h_\mathrm{th}) = \int_{0}^{h_\mathrm{th}} f_h(h) \, dh.
\label{eq:p_out_init}
\end{equation}

By substituting \eqref{eq:pdf_h_meijer} into \eqref{eq:p_out_init} and applying the integration property of the Meijer G-function \cite[Eq. (2.24.2.2)]{Prudnikov-book-1990}, the closed-form expression of $P_\mathrm{out}$ can be obtained as:
\begin{equation}
P_\mathrm{out}(h_\mathrm{th}) = P_\mathrm{int} + (1 - P_\mathrm{int}) \sum_{i=1}^{2} w_i \frac{\Gamma(\alpha_i + \beta_i)}{\Gamma(\alpha_i)} \left(\frac{h_\mathrm{th}}{h_\mathrm{L} A_0}\right)
G_{2,2}^{1,1}\left( \frac{h_\mathrm{th}}{h_\mathrm{L} A_0} \left| \begin{array}{c} 0, \alpha_i + \beta_i - 1 \\ \alpha_i - 1, -1 \end{array} \right.\right),
\label{eq:p_out_final}
\end{equation}
which is strictly valid for $0 \leq h_\mathrm{th} \leq h_\mathrm{L} A_0$.

\begin{figure}[!t]
    \centering
    \includegraphics[width=0.55\textwidth]{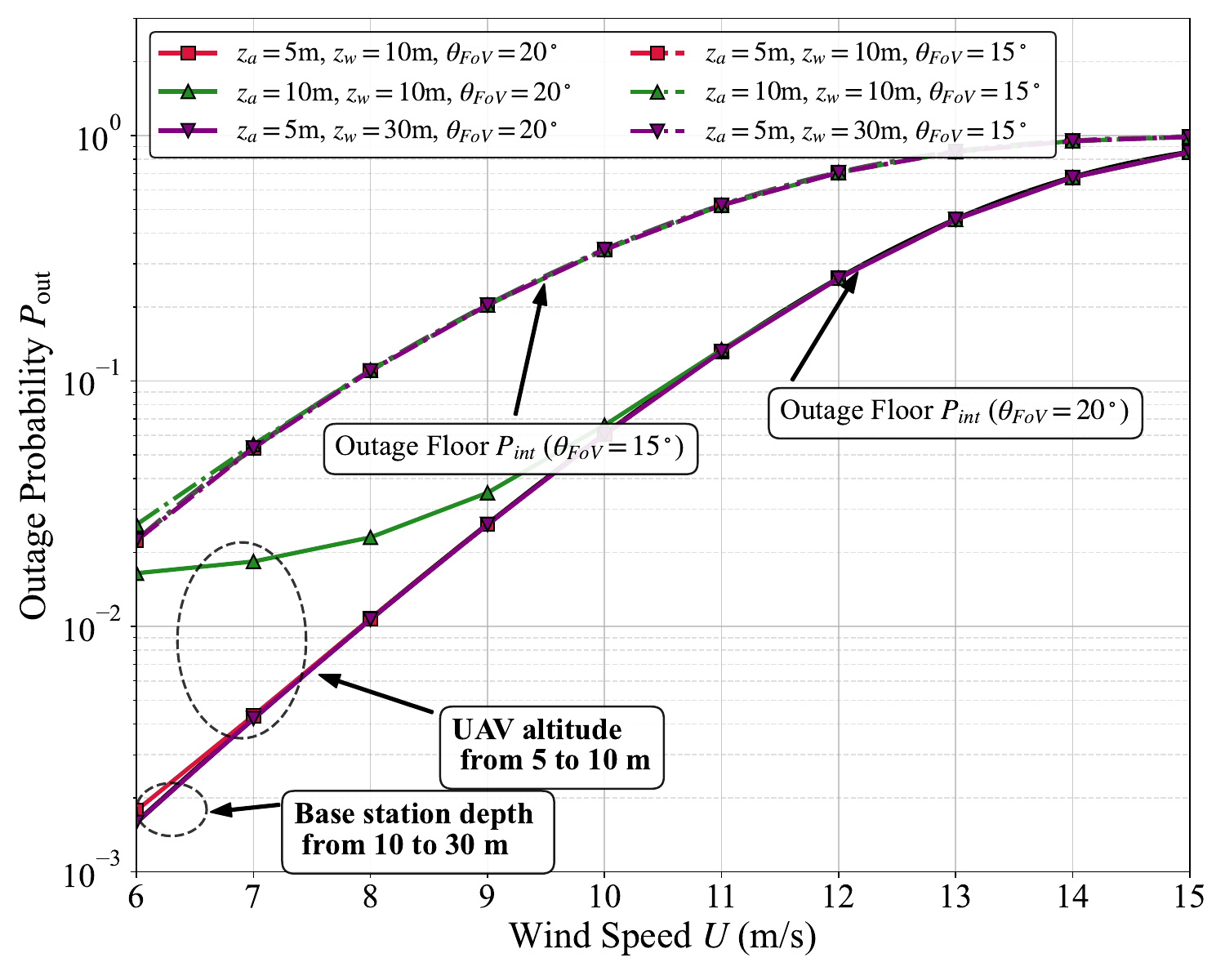}
    \caption{Outage probability versus wind speed for different UAV altitudes ($z_a$) and underwater base station depths ($z_w$).}
    \label{fig:p_out_u}
\end{figure}

Considering a transceiver configuration with an \gls{ld} of half-divergence angle $\theta_0=0.05$\,rad at $\lambda=450$~nm and $P_\mathrm{Tx, 1} = 50$~mW, employing NRZ-OOK signaling with $\xi=0.2$, and a SensL B-series MicroSB 30020 \gls{sipm} at the Rx \cite{Ijeh-JOCN-2022}, along with coastal water conditions characterized by $a(\lambda)=0.0088~\text{m}^{-1}$ and $b(\lambda)=0.216~\text{m}^{-1}$ \cite{Kaushal-Access-2016}, the threshold $h_\mathrm{th}$ is determined considering a target \gls{ber} of $10^{-3}$ using the relation provided in \cite[Eq.~(A4)]{Ijeh-JOCN-2022}.
Figure~\ref{fig:p_out_u} illustrates $P_\mathrm{out}$ versus wind speed for various \gls{tx} depths, \gls{rx} altitudes, and \glspl{fov}, \textcolor{black}{from which three design guidelines can be drawn.} 

First, the outage floor $P_\mathrm{int}$ in~\eqref{eq:pint}, governed by $U$ and $\theta_\mathrm{FoV}$, imposes an irreducible lower bound on $P_\mathrm{out}$ in \eqref{eq:p_out_final}. \textcolor{black}{Since the mean effective \gls{aoa} increases from $8^\circ$ to $26^\circ$, this floor rises sharply and dominates beyond $U \approx 9$~m/s. Widening the \gls{fov} from $15^\circ$ to $20^\circ$ reduces $P_\mathrm{out}$ by an order of magnitude at $U=6$~m/s, yet both configurations saturate beyond $U \approx 14$~m/s. The \gls{fov} should therefore be dimensioned for the worst expected sea state.} \textcolor{black}{Second, below saturation,} increasing the \gls{uav} altitude from $5$ to $10$~m degrades $P_\mathrm{out}$ \textcolor{black}{by nearly one order of magnitude at $U=6$~m/s}, as $r_d$ in \eqref{eq:expr_rd} scales with $z_a$. Lastly, increasing the \gls{tx} depth from $10$ to $30$~m \textcolor{black}{reduces $P_\mathrm{out}$ to the outage floor}: although greater depth increases $h_\mathrm{L}$, the enlarged beam footprint mitigates wave-induced PEs. Owing to the high \gls{sipm} sensitivity, this geometric stabilization outweighs the attenuation penalty.%\textcolor{black}{Link reliability is thus primarily limited by the \gls{fov}, followed by the \gls{uav} altitude, with the \gls{tx} depth providing only a residual margin.}

%%%%%%%%%%%%%%%%%%%%%%%%%%%%%%%%%%%%%%%%%%%%%%%%%%%%%%%%%%%%%%%%%%%%%%%%%%%%%%%%%%%
\vspace{-0.2cm}
\section{Concluding Remarks}
We presented a tractable statistical channel model for vertical \gls{w2a} optical links under dynamic sea-surface refraction \textcolor{black}{and \gls{uav} instability}, characterizing \glspl{pe} via an \gls{em}-fitted \gls{bmm} and deriving exact closed-form $P_\mathrm{out}$ accounting for \gls{tir} and \gls{rx} \gls{fov} limits. Our analysis reveals a fundamental \gls{tir}-induced outage floor that inherently restricts system reliability. It also shows that \gls{uav} altitude degrades performance significantly more than \gls{tx} depth. Ultimately, this framework serves as an efficient tool for designing robust underwater data harvesting systems. \textcolor{black}{Ongoing work aims to analytically relate the \gls{bmm} parameters to the physical link parameters, enabling a fully generalized model for developing system-level mitigation strategies for dynamic \gls{w2a} links.}

\balance
\bibliography{IEEEabrv,biblio}{}

@STRING{IEEE_J_VT         = "{IEEE} Trans. Veh. Technol."}

@STRING{IEEE_J_COM        = "{IEEE} Trans. Commun."}

@STRING{IEEE_J_JLT        = "J. Lightw. Technol."}

@STRING{IEEE_J_SENSOR     = "{IEEE} Sensors J."}

@STRING{IEEE_M_SP         = "{IEEE} Signal Process. Mag."}

@STRING{ELS_LTECH       = "Opt. Laser Technol."}

@STRING{OSA_AM_A        = "J. Opt. Soc. Am. A"}

@STRING{OSA_JOCN        = "{IEEE/OSA} J. Opt. Commun. Netw."}

@article{Ijeh-JOCN-2022,
  title={Outage probability analysis of a vertical underwater wireless optical link subject to oceanic turbulence and pointing errors},
  author={Ijeh, Ikenna Chinazaekpere and Khalighi, Mohammad Ali and Elamassie, Mohammed and Hranilovic, Steve and Uysal, Murat},
  journal=OSA_JOCN,
  volume={14},
  number={6},
  pages={439--453},
  year={2022},
  month={May},
  publisher={Optica Publishing Group}
}

@article{Chen-JOSAA-2024,
  title={Modeling and oblique transmission characteristics of an underwater wireless optical communication channel based on ocean depth layering},
  author={Chen, Dan and Zhao, Peiyan and Tang, Linhai and Wang, Minyan},
  journal=OSA_AM_A,
  volume={41},
  number={3},
  pages={424--434},
  year={2024},
  publisher={Optica Publishing Group},
  month={Feb.}
}

@ARTICLE{Nennouche-Photonics-2026,
  author={Nennouche, Mohamed and Khalighi, Mohammad Ali and Dowhuszko, Alexis Alfredo and Merad, Djamal},
  journal={IEEE Photonics Journal}, 
  title={End-to-End Optical Propagation Modeling for Water-to-Air Channels Under Sea Surface and {UAV} Effects}, 
  year={2026},
  volume={18},
  number={3},
  pages={7301222},
}

@article{Luo-Sensors-2022,
  title={Recent progress of air/water cross-boundary communications for underwater sensor networks: A review},
  author={Luo, Hanjiang and Wang, Jinglong and Bu, Fanfeng and Ruby, Rukhsana and Wu, Kaishun and Guo, Zhongwen},
  journal=IEEE_J_SENSOR,
  volume={22},
  number={9},
  pages={8360--8382},
  year={2022},
  month={Mar.},
  publisher={IEEE}
}

@article{Kaushal-Access-2016,
  title={Underwater optical wireless communication},
  author={Kaushal, Hemani and Kaddoum, Georges},
  journal={{IEEE} Access},
  volume={4},
  pages={1518--1547},
  month={Apr.},
  year={2016},
  publisher={IEEE}
}

@article{Ata-TCOM-2024,
  title={Performance of optical seawater-to-air wireless links in the presence of seawater pitching angle effect},
  author={Ata, Yal{\c{c}}{\i}n and Kiasaleh, Kamran},
  journal=IEEE_J_COM,
  volume={72},
  number={12},
  pages={7856--7865},
  year={2024},
  month={June},
  publisher={IEEE}
}

@article{Rahman-TVT-2022,
  title={Direct air-to-underwater optical wireless communication: Statistical characterization and outage performance},
  author={Rahman, Ziyaur and Zafaruddin, Syed Mohammad and Chaubey, Vinod Kumar},
  journal=IEEE_J_VT,
  volume={72},
  number={2},
  pages={2655--2660},
  year={2022},
  month={Sept.},
  publisher={IEEE}
}

@article{Lin-JLT-2022,
  title={Waving effect characterization for water-to-air optical wireless communication},
  author={Lin, Tianrui and others},
  journal=IEEE_J_JLT,
  volume={41},
  number={1},
  pages={120--136},
  year={2022},
  month={Oct.},
  publisher={IEEE}
}

@article{Angara-OLT-2024,
  title={Influence of sea surface waves and bubbles on the performance of underwater-to-air optical wireless communication system},
  author={Angara, Bhogeswara Rao and Shanmugam, Palanisamy and Ramachandran, Harisankar},
  journal=ELS_LTECH,
  volume={174},
  pages={110652},
  year={2024},
  month={July},
  publisher={Elsevier}
}

@misc{OceanWb-Mobley,
	AUTHOR =       {C. Mobley and E. Boss and C. Roesler},
	TITLE =        {Ocean Optics Web Book},
	address =     {http://www.oceanopticsbook.info/},
	note = {last accessed: 26 Mar. 2026},
}

@inproceedings{Ijeh-CSNDSP-2026,
      title={Air-Sea Surface Modeling and Operating Link Range Evaluation for {AUV-to-UAV} Optical Wireless Communication Links}, 
      author={Ikenna Chinazaekpere Ijeh and Mohammad Ali Khalighi and Wasiu O. Popoola},
  booktitle={Int. Symp. Commun. Syst., Netw. Dig. Sig. Proc. (CSNDSP)},
      year={2026},
        pages={1-6},
}

@article{Moon-SPM-1996,
  title={The expectation-maximization algorithm},
  author={Moon, Todd K},
  journal=IEEE_M_SP,
  volume={13},
  number={6},
  pages={47--60},
  year={1996},
  month={Nov.},
  publisher={IEEE}
}

@book{Prudnikov-book-1990,
  title={Integrals and Series, volume 3: more special functions},
  author={Prudnikov, AP and Brychkov, Yu A and Marichev, OI and others},
  volume={3},
  year={1990},
  publisher={Gordon and Breach Science Publishers}
}

@article{Wang-TVT-2024,
  title={Enabling reliable water--air direct optical wireless communication for uncrewed vehicular networks: A deep reinforcement learning approach},
  author={Wang, Jinglong and Luo, Hanjiang and Ruby, Rukhsana and Liu, Jiangang and Wang, Shengli and Wu, Kaishun},
  journal=IEEE_J_VT,
  volume={73},
  number={8},
  pages={11470--11486},
  year={2024},
  month={Mar.},
  publisher={IEEE}
}
\bibliographystyle{IEEEtran}
\balance

\end{document}